\documentclass[prl,twocolumn,superscriptaddress,nofootinbib,
aps,10pt,preprintnumbers]{revtex4-1}

\usepackage[colorlinks=true,linkcolor=black,citecolor=blue,urlcolor=blue, pdfborder={0 0 0}]{hyperref}
\usepackage{mathtools}
\usepackage[utf8]{inputenc}
\usepackage{color}
\usepackage[table,xcdraw,dvipsnames]{xcolor}
\usepackage{subfigure}
\usepackage{amssymb}
\usepackage[normalem]{ulem}
\usepackage{slashed}

\newcommand{ \mysmall}[1]{\scriptscriptstyle #1} 

\newcommand{\Sec}[1]{ \medskip \noindent {\sl \bfseries #1}}

\def\eq#1{{Eq.~(\ref{#1})}}
\def\eqs#1#2{{Eqs.~(\ref{#1})--(\ref{#2})}}
\def\fig#1{{Fig.~\ref{#1}}}

\def\abs#1{\left| #1\right|}

\renewcommand{\bar}{\overline}

\newcommand{\beq}{\begin{equation}}
\newcommand{\eeq}{\end{equation}}
\newcommand{\bea}{\begin{eqnarray}}
\newcommand{\eea}{\end{eqnarray}}

\renewcommand{\(}{\left(}
\renewcommand{\)}{\right)}

\renewcommand{\(}{\left(}
\renewcommand{\)}{\right)}

\definecolor{Gray}{gray}{0.95}
\definecolor{RGray}{gray}{0.85}
\definecolor{CGray}{gray}{0.93}
     
\begin{document}

\title{New physics behind the new muon \boldmath{$g$}-2 puzzle?}

\newcommand{\affPD}{{\small \it Dipartimento di Fisica e Astronomia `G.~Galilei', Universit\`a di Padova, Italy 
}}

\newcommand{\affPDINFN}{{\small \it Istituto Nazionale di Fisica Nucleare, Sezione di Padova, Padova, Italy}}

\author{Luca Di Luzio}
\affiliation{\affPD}
\affiliation{\affPDINFN}

\author{Antonio Masiero}
\affiliation{\affPD}
\affiliation{\affPDINFN}

\author{Paride Paradisi}
\affiliation{\affPD}
\affiliation{\affPDINFN}

\author{Massimo Passera}
\affiliation{\affPDINFN}

\begin{abstract}
The recent measurement of the muon $g$-2 at Fermilab confirms the previous Brookhaven result.  
The leading hadronic vacuum polarization (HVP) contribution to the muon $g$-2 represents a crucial 
ingredient to establish if the Standard Model prediction differs from the experimental value. 
A recent lattice QCD result by the BMW collaboration shows a tension with the low-energy 
$e^+e^- \to \text{hadrons}$ data which are currently used to determine the HVP contribution. 
We refer to this tension as the new muon $g$-2 puzzle. 
In this Letter we consider  the possibility that new physics 
contributes to the 
$e^+e^- \to \text{hadrons}$ cross-section. 
This scenario could, in principle, solve the new muon $g$-2 puzzle. 
However, we show that this solution is excluded by a number of experimental constraints. 

\end{abstract}

\maketitle

\Sec{Introduction.} 
The anomalous magnetic moment of the muon, 
$a_\mu \!\equiv\! (g_\mu \!-\! 2)/2$, 
has provided a persisting hint of new physics (NP) for many years. 
The recent $a_\mu$ measurement by the Muon $g$-2 collaboration at Fermilab~\cite{Muong-2:2021ojo,Muong-2:2021vma,Muong-2:2021ovs,Muong-2:2021xzz} 
has confirmed the earlier 
result 
by the E821 experiment at Brookhaven~\cite{Muong-2:2006rrc}, yielding the average 
$a_\mu^{\mysmall \rm EXP} \!=\! 116592061(41) \!\times\! 10^{-11}$. 
The comparison of this result with the Standard Model (SM) prediction $a_\mu^{\mysmall \rm SM} \!=\! 116591810(43) \times 10^{-11}$
of the Muon $g$-2 Theory Initiative~\cite{Aoyama:2020ynm} leads to an intriguing $4.2\sigma$ discrepancy~\cite{Muong-2:2021ojo}
\begin{equation}
\Delta a_\mu = a_\mu^{\mysmall \rm EXP}-a_\mu^{\mysmall \rm SM} = 251 \, (59) \times 10^{-11}\,.
\label{eq:gmu}
\end{equation}
The expected forthcoming results of the Fermilab 
experiment plan to reach a sensitivity four-times better than the E821 one. 
Moreover, in a longer term, 
also the E34 collaboration at J-PARC~\cite{Abe:2019thb} aims at measuring the muon $g$-2 through a new low-energy approach. 

\begin{figure}[ht]
\centering
\includegraphics[height=2.7cm]{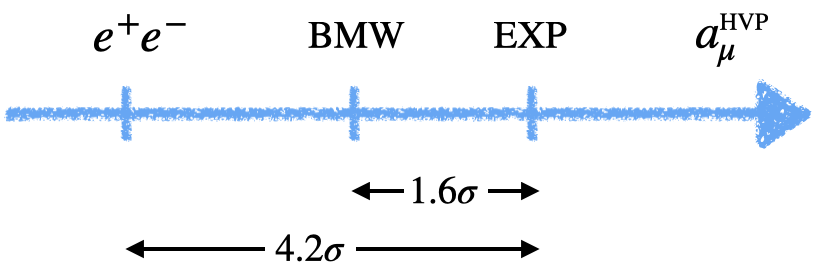} 
\caption{The new muon $g$-2 puzzle: $4.2 \sigma$ vs.~$1.6 \sigma$.}
\label{fig:newgm2puzzle}
\end{figure}

On the theory side, the only source of sizable uncertainties in $a_\mu^{\mysmall \rm SM}$ stems from the non-perturbative contributions of the hadronic sector,
which have been under close scrutiny for several years.
The SM prediction $a_\mu^{\mysmall \rm SM}$ in Eq.~(\ref{eq:gmu}) has been derived using 
$(a_\mu^{\rm \mysmall HVP})^{\rm \mysmall TI}_{e^+e^-}$, 
the leading hadronic vacuum polarization (HVP) contribution to the muon $g$-2 
based on low-energy $e^+e^- \!\to\! {\rm hadrons}$ data 
obtained by 
the Muon $g$-2 Theory Initiative \cite{Aoyama:2020ynm}
(see also \cite{Jegerlehner:2017gek,Davier:2017zfy,Keshavarzi:2018mgv,Colangelo:2018mtw,Hoferichter:2019mqg,Davier:2019can,Keshavarzi:2019abf,Hoid:2020xjs,Kurz:2014wya,Melnikov:2003xd,Masjuan:2017tvw,Colangelo:2017fiz,Hoferichter:2018kwz,Gerardin:2019vio,Bijnens:2019ghy,Colangelo:2019uex,Blum:2019ugy,Aoyama:2012wk,Aoyama:2019ryr,Czarnecki:2002nt,Gnendiger:2013pva}).
Alternatively, the HVP contribution has been computed using a first-principle lattice QCD approach \cite{Aoyama:2020ynm} 
(see also \cite{Budapest-Marseille-Wuppertal:2017okr,RBC:2018dos,Giusti:2019xct,FermilabLattice:2019ugu,Gerardin:2019rua}).
Recently, the BMW lattice QCD collaboration (BMWc) computed 
the leading HVP contribution to the muon $g$-2 with sub per-cent precision,
finding a value, $(a_\mu^{\rm \mysmall HVP})_{\rm \mysmall BMW}$, larger than $(a_\mu^{\rm \mysmall HVP})_{e^+e^-}^{\rm \mysmall TI}$~\cite{Borsanyi:2020mff}.
If $(a_\mu^{\rm \mysmall HVP})_{\rm \mysmall BMW}$ is used to obtain $a_\mu^{\mysmall \rm SM}$ 
instead of $(a_\mu^{\rm \mysmall HVP})_{e^+e^-}^{\rm \mysmall TI}$, the discrepancy with the experimental result is 
reduced to $1.6\sigma$ only. 
The above results are respectively 
\begin{align} 
\label{eq:currentB}
&(a_\mu^{\rm \mysmall HVP})_{e^+e^-}^{\rm \mysmall TI} = 6931\, (40) \times 10^{-11}\, ,
\\
\label{eq:currentA}
& (a_\mu^{\rm \mysmall HVP})_{\rm \mysmall BMW} = 7075\, (55) \times 10^{-11}\, .
\end{align}
The present situation regarding the leading HVP contribution to the muon $g$-2 can be schematically represented as in \fig{fig:newgm2puzzle},
where $(a_\mu^{\rm \mysmall HVP})_{\mysmall \rm EXP}$ is the value of the HVP contribution required to exactly match $a_\mu^{\mysmall \rm EXP}$ assuming no NP.
Hereafter, the difference between the discrepancies in \fig{fig:newgm2puzzle} will be referred to as the {\it new muon $g$-2 puzzle}.

Assuming that both 
$(a_\mu^{\rm \mysmall HVP})_{e^+e^-}^{\rm \mysmall TI}$ and $(a_\mu^{\rm \mysmall HVP})_{\rm \mysmall BMW}$ are correct, 
we ask whether this puzzle can be solved thanks to NP effects 
which would bring $(a_\mu^{\rm \mysmall HVP})_{e^+e^-}^{\rm \mysmall TI}$ in agreement with 
$(a_\mu^{\rm \mysmall HVP})_{\rm \mysmall BMW}$, 
\emph{without} spoiling the $1.6\sigma$ agreement of $(a_\mu^{\rm \mysmall HVP})_{\rm \mysmall BMW}$ with 
$(a_\mu^{\rm \mysmall HVP})_{\mysmall \rm EXP}$.  
Differently from what has been usually done in the literature, here we do not assume a direct NP contribution to $\Delta a_\mu$   
(i.e.~new states that couple directly to muons). If fact, by itself this possibility could solve the longstanding discrepancy in \eq{eq:gmu}, 
but not the new muon $g$-2 puzzle. Instead, in order to solve the latter,
we invoke NP that modifies the $e^+e^- \!\to\! {\rm hadrons}$ cross-section $\sigma_{\rm had}$.

\begin{figure*}[th]
\centering
\hspace{-0.3cm}
\includegraphics[height=2.1cm]{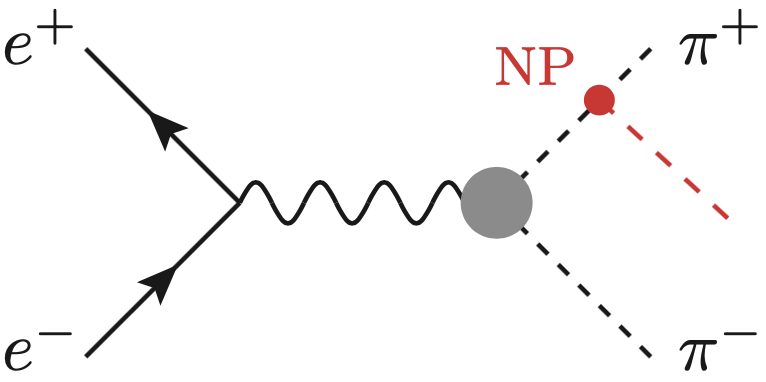} \ \ \ \ \
\includegraphics[height=2.1cm]{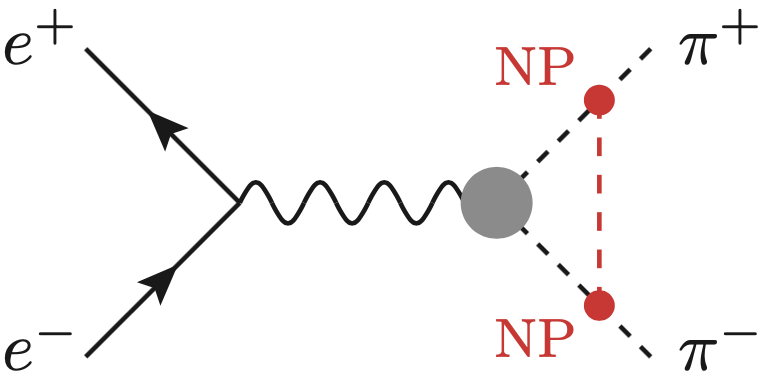} \ \ \ \
\includegraphics[height=2.1cm]{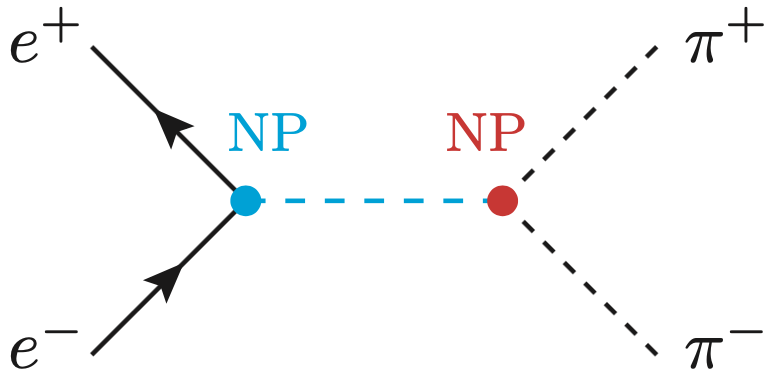}  
\caption{
Examples of NP contributions to $\sigma_{\rm had}$ via FSR (first and second diagram) 
and via a NP tree-level mediator coupled both to hadrons and electrons 
(third diagram). 
}
\label{fig:NPinVP}
\end{figure*}

An increase of $\sigma_{\rm had}$, 
due to an unforeseen missing contribution,
has been already proposed to enhance $(a_\mu^{\rm \mysmall HVP})_{e^+e^-}^{\rm \mysmall TI}$ 
and solve 
$\Delta a_\mu$
\cite{Passera:2008jk,Crivellin:2020zul,Keshavarzi:2020bfy,Malaescu:2020zuc,Colangelo:2020lcg}.
However, the required shift in $\sigma_{\rm had}$ is disfavoured by the electroweak fit if it occurs at $\sqrt{s}\gtrsim 1$ GeV \cite{Keshavarzi:2020bfy}. 
Hence, in the following, we will consider NP modifications of $\sigma_{\rm had}$ below the GeV scale.
While Refs.~\cite{Passera:2008jk,Crivellin:2020zul,Keshavarzi:2020bfy,Malaescu:2020zuc,Colangelo:2020lcg} did not specify the origin of the shift in $\sigma_{\rm had}$, we here assume that it is due to NP. 
After classifying in a model-independent way 
the general properties of such a NP, we investigate for the first time 
its non-trivial impact on $e^+e^-$ and BMWc lattice results.  

Crucial for our analysis is the dispersion relation used to determine $(a_\mu^{\rm \mysmall HVP})_{e^+e^-}^{\rm \mysmall TI}$ via  
$\sigma_{\rm had}$. 
This relation follows from the application of the optical theorem to the photon HVP. 
It will be shown that 
scenarios in which NP couples only to hadrons 
are not able to solve the new muon $g$-2 puzzle.  
Instead, if NP 
couples both to hadrons and electrons 
(and hence it contributes to $\sigma_{\rm had}$ at tree level),
$(a_\mu^{\rm \mysmall HVP})_{e^+e^-}^{\rm \mysmall TI}$ 
should be subtracted of NP 
in the comparison with $(a_\mu^{\rm \mysmall HVP})_{\mysmall \rm EXP}$, whereas $(a_\mu^{\rm \mysmall HVP})_{\rm \mysmall BMW}$ should not. 
In fact, in this case, 
the quantity that should enter the dispersion relation determining 
the HVP contribution 
should not be the experimentally measured cross-section $\sigma_{\rm had}$, 
but rather $\sigma_{\rm had} - \Delta\sigma^{\rm\mysmall NP}_{\rm had}$. 
Therefore, the tension between $(a_\mu^{\rm \mysmall HVP})_{e^+e^-}^{\rm \mysmall TI}$ and $(a_\mu^{\rm \mysmall HVP})_{\rm \mysmall BMW}$ 
could be solved by invoking a negative interference between the SM and NP, that is $\Delta\sigma^{\rm\mysmall NP}_{\rm had}<0$.
As we will show below, the above picture selects a very specific NP scenario which entails new light particles with a mass scale $\lesssim 1$ GeV 
coupling to SM fermions through a vector current.

\Sec{Model-independent analysis.}
Hereafter we are going to examine the general properties of NP models aiming at solving the new muon $g$-2 puzzle 
via a modification of $\sigma_{\rm had}$. 
To this end, we introduce the dispersion relation 
\begin{align} 
\label{eq:sigmatoa}
(a_\mu^{\rm \mysmall HVP})_{e^+e^-} &=
\frac{\alpha}{\pi^2}
\int_{m^2_{\pi^0}}^{\infty} \frac{\text{d}s}{s} \, K(s) \, {\rm Im} \, \Pi_{\rm had} (s) \, , 
\end{align}
where $K(s)$ is a positive-definite kernel function with $K(s) \approx m^2_\mu / 3s$ for $\sqrt{s} \gg m_\mu$. 
This equation defines the HVP contribution to the muon $g$-2 in terms of the photon HVP, $\Pi_{\rm had}$, 
which includes possible NP effects.
If the possible NP entering the photon HVP does not couple to electrons, 
i.e.~it does not enter the hadronic cross-section at tree level, 
then Eq.~(\ref{eq:sigmatoa}) can be written as 
\begin{align} 
\label{eq:sigmatoab}
(a_\mu^{\rm \mysmall HVP})_{e^+e^-} &= \frac{1}{4\pi^3} \int_{m^2_{\pi^0}}^{\infty} 
\text{d}s \, K(s) \sigma_{\rm had}(s) \,, 
\end{align}
where   
$\sigma_{\rm had}$ 
includes final-state radiation (FSR), whereas both vacuum polarization and initial-state radiation (ISR) effects are subtracted. 
In particular, vacuum polarization corrections can be simply accounted for by multiplying the experimental cross-section by $|\alpha/\alpha(s)|^2$, while the correction of ISR and 
ISR/FSR interference effects is addressed by each experimental collaboration. 
In this Letter we will focus on the region where $\sigma_{\rm had}$ is experimentally determined, i.e.~$\sqrt{s} \gtrsim 0.3$ GeV, 
since this gives the by far dominant contribution to the dispersive integral in \eq{eq:sigmatoab}. 

In \fig{fig:NPinVP} we show a schematic classification of how NP can enter $\sigma_{\rm had}$. 
The first two diagrams are representative of FSR effects,  
which also 
unavoidably affect the photon HVP at the next-to-leading order (NLO).
We can safely neglect possible NP contaminations in ISR since the bounds on NP couplings to electrons are very severe. 
The third diagram, where NP enters the hadronic cross-section at tree level 
coupling both to hadrons and electrons, is due to NP that also modifies the photon HVP at NLO.  
Crucially, however, its dominant contribution to the muon $g$-2 emerges via the tree-level shift of $\sigma_{\rm had}$.  

Hence, when invoking NP in $\sigma_{\rm had}$, there are two different scenarios to be considered, 
depending on whether 
NP couples only to hadrons or both to hadrons and electrons.
In the following, we analyze these two possibilities and their capability to solve the new muon $g$-2 puzzle.

\smallskip
{\it{1.\,\,NP coupled only to hadrons}}.
This scenario is schematically represented by the first two diagrams of \fig{fig:NPinVP}.
As remarked above, real and virtual FSR must be included in $\sigma_{\rm had}$.
However, in order to establish the impact of NP in FSR (which depends on the 
interplay between the mass scale of NP and the experimental cuts), it would be mandatory to perform 
dedicated experimental analyses imposing the various selection cuts specific of each experimental setup, which is beyond the scope of this Letter. 
Since the full photon FSR effect estimated in scalar QED amounts only to $50 \times 10^{-11}$ \cite{Aoyama:2020ynm,Schwinger:1989ix}, 
a value well below the discrepancy between \eqs{eq:currentB}{eq:currentA}, 
and given that light NP couplings with the SM particles are tightly constrained, the NP contributions in FSR 
cannot solve the new muon $g$-2 puzzle.

\smallskip

{\it{2.\,\,NP coupled both to hadrons and electrons}}.
If NP contributes to $\sigma_{\rm had}$ at tree level
(see third diagram in \fig{fig:NPinVP}), then only 
the subtracted cross-section
$\sigma_{\rm had} - \Delta \sigma^{\rm \mysmall NP}_{\rm had}$
should be included in \eq{eq:sigmatoab}. We note that the latter can be larger than 
$\sigma_{\rm had}$ if $\Delta \sigma^{\rm \mysmall NP}_{\rm had} < 0$, 
thus requiring that the NP contribution is dominated by a \emph{negative} interference with the SM. 
As $(a_\mu^{\rm \mysmall HVP})_{e^+e^-}^{\rm \mysmall TI}$ has been computed 
using $\sigma_{\rm had}$ rather than the subtracted cross-section $\sigma_{\rm had} - \Delta \sigma^{\rm \mysmall NP}_{\rm had}$, 
the theoretical prediction of the HVP contribution in \eq{eq:sigmatoab} is
\begin{align}
\label{eq:noVP_exp}
(a_\mu^{\rm \mysmall HVP})_{e^+e^-} = (a_\mu^{\rm \mysmall HVP})_{e^+e^-}^{\rm \mysmall TI} + (a_\mu^{\rm \mysmall HVP})_{\rm \mysmall NP} \, ,  
\end{align} 
where $(a_\mu^{\rm \mysmall HVP})_{\rm \mysmall NP}$ describes NP 
effects at LO, due to the tree-level exchange of the NP mediator (see third diagram in \fig{fig:NPinVP}), as well as at NLO. 
Instead, 
$(a_\mu^{\rm \mysmall HVP})_{\rm \mysmall BMW}$ should be shifted only by NLO NP effects. 
Remarkably, this scenario may allow to match Eq.~(\ref{eq:noVP_exp}) with $(a_\mu^{\rm \mysmall HVP})_{\rm \mysmall EXP}$, 
while keeping at the same time the agreement with the BMWc lattice result. 

\Sec{Light new physics analysis.}
We now explore whether the second scenario envisaged above can be quantitatively realized in an explicit NP model. 
Motivated by the fact that the kernel function in \eq{eq:sigmatoab} scales like $1/s$ and by the fact that modifications 
of $\sigma_{\rm had}$ above $\sim 1$ GeV are disfavoured by electroweak precision tests \cite{Keshavarzi:2020bfy}, 
we focus on the sub-GeV energy range, where the dominant contribution to $\sigma_{\rm had}$ 
arises from the $e^+e^- \to \pi^+\pi^-$ channel. In fact, in the SM, this channel 
accounts for more than $70\%$ of the full hadronic contribution to the muon $g$-2. 
Furthermore, the requirement of having a sizeable negative interference with the SM amplitude narrows down the general class of NP models. 
Indeed, the interference of scalar couplings with the SM vector current is suppressed by the electron mass, while pseudoscalar and axial couplings 
do not interfere. 
Hence, we focus on the tree-level exchange of a light $Z'$ boson with the following vector couplings to electrons and first-generation 
quarks\footnote{Assuming only the interactions in \eq{eq:ZpLag}, the gauge current associated to the $Z'$ would be anomalous. 
Additional heavy fermions charged under the electroweak group can be introduced in order to make the model UV consistent.}
\beq 
\label{eq:ZpLag}
\mathcal{L}_{Z'} \supset 
(g_{V}^{e} \, \bar e \gamma^\mu e + g_{V}^{q} \, \bar q \gamma^\mu q) Z'_\mu \, ,
\eeq
with $q=u,d$ and $m_{Z'} \lesssim 1$ GeV. 

The matrix element of the two-pion intermediate state can be expressed in terms of the pion vector form factor 
\beq 
\label{eq:pipiVFF} 
\langle \pi^{\pm} (p') | 
J_{\rm em}^\mu(0) | 
\pi^{\pm} (p) \rangle = 
\pm (p'+p)^\mu F^V_\pi(q^2) \, ,
\eeq
where $q = p'-p$ and 
$J_{\rm em}^\mu = \frac{2}{3} \bar u \gamma^\mu u - \frac{1}{3} \bar d \gamma^\mu d$ is the electromagnetic current. 
Exploiting the charge conjugation invariance, in the iso-spin symmetric limit, 
we find that
\beq 
\label{eq:isospinJ}
\langle \pi^{\pm} | 
J_{\rm em}^\mu | 
\pi^{\pm} \rangle 
= 
\langle \pi^{\pm} | 
\bar u \gamma^\mu u | 
\pi^{\pm} \rangle
= - 
\langle \pi^{\pm} | 
\bar d \gamma^\mu d | 
\pi^{\pm} \rangle \, . 
\eeq 
Hence, $F^V_\pi(q^2)$ also describes the matrix element of the $Z'$ quark current 
$J^\mu_{Z'} = g^u_V \bar u \gamma^\mu u + g^d_V \bar d \gamma^\mu d$ 
\begin{align} 
\label{eq:pipiVFFZp} 
\!\!\!\!\langle \pi^{\pm}\! (p') | 
J_{Z'}^\mu(0) | \pi^{\pm} (p) \rangle &\!= \pm (p'+p)^\mu F^V_\pi \! (q^2) (g^u_V \!-\! g^d_V). 
\end{align}
Note that if the $Z'$ interactions are iso-spin symmetric, then $g_V^u = g_V^d$ and the contribution to the $\pi^+\pi^-$ amplitude vanishes. 
Defining $\sigma_{\pi\pi}^{\rm \mysmall SM + NP}= \sigma_{\pi\pi}^{\rm \mysmall SM} + \Delta\sigma_{\pi\pi}^{\rm \mysmall NP}$, 
the tree-level exchange of the $Z'$ 
with width $\Gamma_{Z'}$ 
leads to
\begin{align} 
\label{eq:SMNPoSMeepipi}
\!\!\!\frac{\sigma_{\pi\pi}^{\rm \mysmall SM + NP}}{\sigma_{\pi\pi}^{\rm \mysmall SM}} 
&= \abs{1 + \frac{g_V^e (g_V^u - g_V^d)}{e^2} \frac{s}{s- m^2_{Z'} + i m_{Z'} \Gamma_{Z'}}}^2 ,
\end{align} 
where 
the pion vector form factor 
cancels out in the ratio.

The dispersive contribution to the muon $g$-2 due to SM and NP can be obtained by using 
$\sigma_{\rm had} - \Delta \sigma^{\rm \mysmall NP}_{\rm had}$ in \eq{eq:sigmatoab}. 
Imposing that the current discrepancy $\Delta a_\mu$ is solved by NP in the hadronic cross-section, we obtain 
\beq
\label{eq:Deltaamupointl} 
\Delta a_\mu =  \frac{1}{4\pi^3} \int_{s_{\rm exp}}^{\infty} \text{d}s\, K(s) (- \Delta\sigma^{\rm \mysmall NP}_{\rm had}(s)) \, , 
\eeq
where the lower integration limit is $s_{\rm exp} \approx (0.3 \ \text{GeV})^2$, that is, the 
integral is performed in the
data-driven region for the $\pi\pi$ channel.  
Approximating $\Delta\sigma^{\rm \mysmall NP}_{\rm had} \approx \Delta \sigma_{\pi\pi}^{\rm \mysmall NP}$,\footnote{We expect that 
this approximation reproduces $\Delta a_\mu$ with $\mathcal{O}(20)\%$ accuracy. Indeed, in the SM, the $\pi^0 \gamma$ and $n \pi$ 
channels (with $n>2$) amount to 17$\%$ of the $\pi\pi$ channel contribution to $(a_\mu^{\rm \mysmall HVP})_{e^+e^-}^{\rm \mysmall TI}$ 
(see e.g.~\cite{Keshavarzi:2019abf}). Moreover, NP is assumed to couple only to up and down quarks so that the contribution of other heavy mesons is negligible.} 
from \eq{eq:SMNPoSMeepipi} we find
\begin{align}
\label{eq:Deltasigma}
\Delta\sigma^{\rm \mysmall NP}_{\rm had} (s)
&\approx \sigma_{\pi\pi}^{\rm \mysmall SM} (s) \times \frac{ 2 \epsilon s (s- m^2_{Z'}) + \epsilon^2 s^2}{(s- m^2_{Z'})^2 + m_{Z'}^4\gamma^2} \, , 
\end{align} 
where we introduced the effective coupling 
$\epsilon \equiv g_V^e (g_V^u - g_V^d) / e^2$ 
and the adimensional width parameter 
$\gamma \equiv \Gamma_{Z'} / m_{Z'}$.  
If both the $Z' \to e e$ and $Z' \to \pi^+ \pi^-$ channels are kinematically open,  
the associated decay widths (normalized to $m_{Z'}$) read, respectively 
\beq 
\label{eq:gammaee}
\gamma_{ee} \approx \frac{(g_V^e)^2}{12\pi} = 2.7 \times 10^{-10} \(\frac{g_V^e}{10^{-4}} \)^2 \, ,
\eeq
up to $(m_e / m_{Z'})^4$ corrections, and
\beq
\label{eq:gammapipi}
\gamma_{\pi\pi} = \frac{(g_V^u - g_V^d)^2}{48 \pi} |F^V_\pi (m^2_{Z'})|^2 \( 1 - \frac{4m^2_\pi}{m^2_{Z'}} \)^{3/2} \, , 
\eeq
where $|F^V_\pi (m_{Z'}^2)|^2$ (normalized to $F^V_\pi (0)=1$) can be enhanced 
up to a factor of 45 by the $\rho$ resonance \cite{Colangelo:2018mtw}. 
For $m_{Z'} < 2 m_{\pi^+} \approx 0.28$ GeV, $\gamma = \gamma_{ee}$, whereas for $m_{Z'} \in [0.3, 1]$ GeV we can approximate 
$\gamma \approx \gamma_{\pi\pi}$ since the $e^+e^-$ channel can be safely neglected given the tight bounds on 
$g^e_V$. 
Note that possible contributions to $\gamma$ stemming from non-SM final states (e.g.~decays into a dark sector) 
yield a positive-definite shift to $\sigma_{\rm had}$, since they cannot interfere with the SM. Hence, they contribute with a negative shift to $\Delta a_\mu$ (cf.~\eq{eq:Deltaamupointl}), 
thus worsening the discrepancy. 

\begin{figure}[ht!]
\centering
\includegraphics[height=7.5cm]{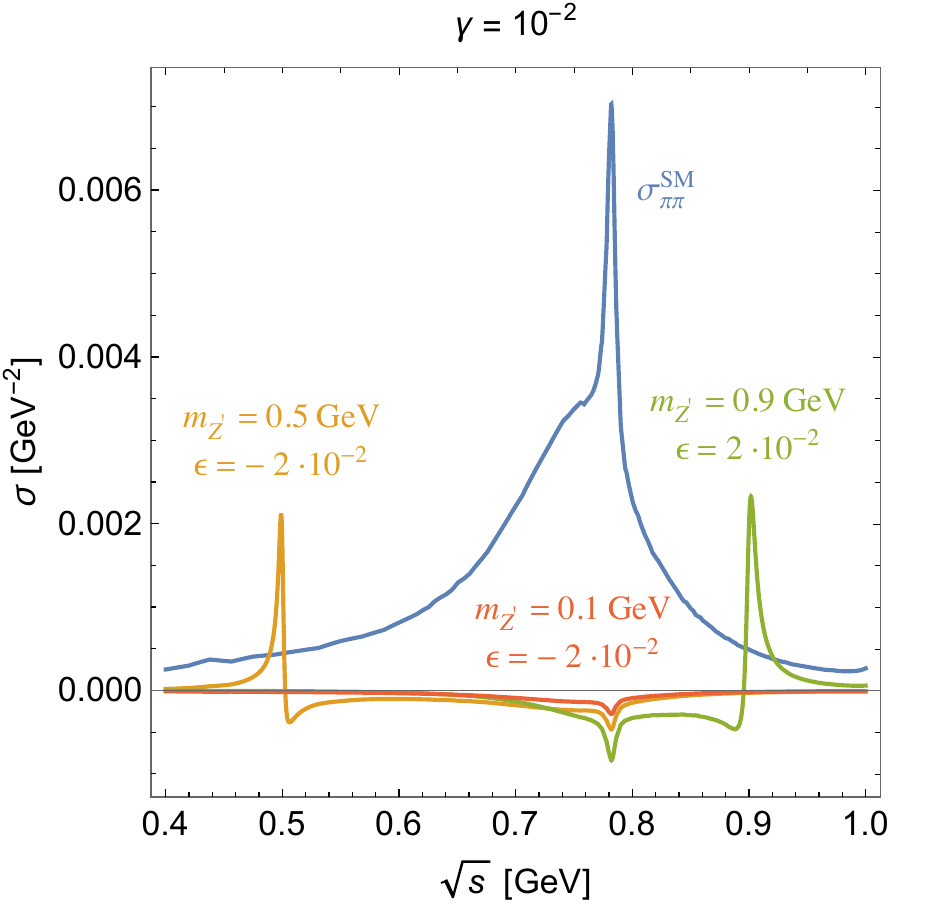}
\caption{$\sigma^{\rm \mysmall SM}_{\rm had}$ and $\Delta\sigma^{\rm \mysmall NP}_{\rm had}$   
for some benchmark values of the $Z'$ model parameters solving the $\Delta a_\mu$ discrepancy, see Eq.~\eqref{eq:gmu}.} 
\label{fig:sigmaSMandNP}
\end{figure}

\begin{figure*}[ht]
\centering
\includegraphics[height=7.5cm]{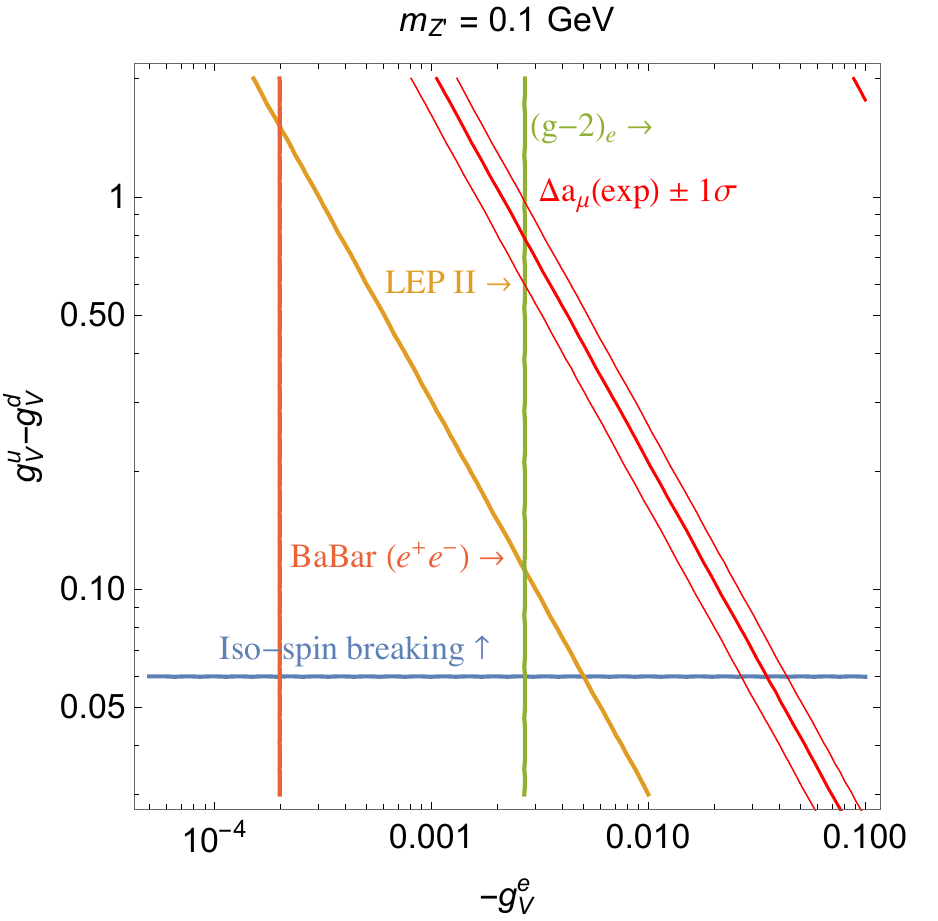} \qquad 
\includegraphics[height=7.5cm]{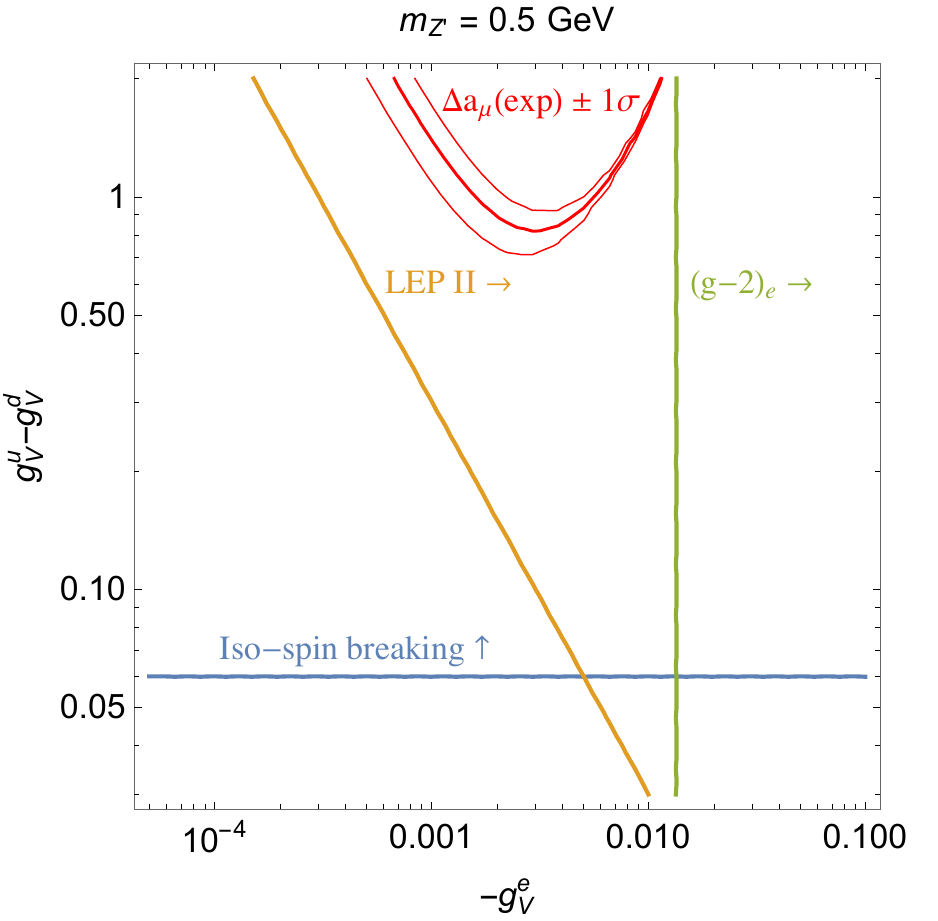}
\caption{$Z'$ contribution to $\Delta a_\mu$ via a modification of $\sigma_{\rm had}$ vs.~$Z'$ constraints.}
\label{fig:gegudmZp}
\end{figure*}

The profile of $\Delta\sigma^{\rm \mysmall NP}_{\rm had}$, together with its SM counterpart (taken from Ref.~\citep{Keshavarzi:2019abf}), 
is shown in \fig{fig:sigmaSMandNP} for some representative benchmark values of the $Z'$ model parameters addressing the 
$\Delta a_\mu$ discrepancy.  
In particular, we find that 
$Z'$ masses below the $\rho$ resonance require $\epsilon < 0$, 
whereas $Z'$ masses above it require $\epsilon > 0$. 
Moreover, in order to obtain $\Delta a_\mu > 0$, the interference term has to dominate over the pure resonant effect in Eq.~(\ref{eq:Deltasigma}).
All in all, we find that the parameters of the $Z'$ model that are needed to explain $\Delta a_\mu$ can be divided in two regions: 
$i)$ $m_{Z'} \gtrsim 0.3$ GeV which requires $|\epsilon| \approx 10^{-2}$ and $\gamma \gtrsim 10^{-3}$ and $ii)$ $m_{Z'} \lesssim 0.3$ GeV 
which requires $|\epsilon| \approx 10^{-2}$ and basically no relevant constraint on $\gamma$ 
(as evident from \eq{eq:Deltasigma}). 
We note that in principle 
it could be possible to directly observe (with a dedicated scanning analysis) the new resonance in $e^+e^-$ data 
for particular choices of the $Z'$ mass and width parameters. 
However, 
since there are also configurations that can explain $\Delta a_\mu$ with a very narrow and light $Z'$ 
below the experimental resolution, it still remains necessary to constrain the $Z'$ couplings indirectly. 

In the following, we are going to inspect whether the region of the parameter space of the $Z'$ model needed to explain 
$\Delta a_\mu$ is allowed by 
experimental constraints. These can be divided for convenience in three classes: 
1.~semi-leptonic processes; 2.~purely leptonic processes 
and 3.~purely hadronic, iso-spin violating observables.

\smallskip
{\it 1.~Semi-leptonic processes.} 
The $e^+e^- \to q \bar q$ cross-section, $\sigma_{qq}$, has been measured with a per-cent level accuracy at LEP II for center of mass energies 
$\sqrt{s} \in [130, 207]$ GeV~\cite{ALEPH:2013dgf}. The leading effect to this process induced by a $Z'$ exchange is given by
\begin{align} 
\label{eq:SMNPoSMeeqq}
\frac{\sigma_{qq}^{\rm\mysmall SM + NP}}{\sigma_{qq}^{\rm\mysmall SM}} 
&\approx 1 + 2 \frac{g_V^e g_V^q}{e^2 Q_q} 
\, , 
\end{align} 
where $Q_q$ denotes the quark charge.
Requiring that the deviation from unity in \eq{eq:SMNPoSMeeqq} is less than 1$\%$~\cite{ALEPH:2013dgf} 
leads to 
$|g_V^e g_V^q| \lesssim 4.6 \cdot 10^{-4} |Q_q|$
which translates into 
$\epsilon \lesssim 3.3 \cdot 10^{-3}$. 
Moreover, 
the bound does not depend on the $Z'$ mass 
and it acts on a coupling combination that is 
similar to the one entering $\Delta\sigma_{\pi\pi}^{\rm \mysmall NP}$, 
but not vanishing in the iso-spin symmetric limit 
$g_V^u = g_V^d$. Hence, this bound can be considered to be conservative. 

\smallskip
{\it 2.~Leptonic processes.} 
The $Z'$ coupling to electrons is also tightly constrained. In particular, the non-observation at BaBar of the process $e^+e^- \to \gamma Z'$ 
followed by the decay $Z' \to e^+e^-$ yields $g^e_V \lesssim 2 \cdot 10^{-4}$~\cite{BaBar:2014zli} if the ${Z'}$ decays dominantly into electrons.
Therefore, in our framework, this bound applies only for $m_{Z'} \lesssim 0.3$ GeV where the $Z'\to \pi^+\pi^-$ decay is not kinematically allowed.
Moreover, another important bound arises from the electron $g$-2, yielding  
$|g_V^e| \lesssim 10^{-2}\, (m_{Z'}/0.5$ GeV) for $m_{Z'} \gtrsim$ MeV. 

\smallskip
{\it 3.~Iso-spin breaking observables.} 
The $Z'$ contribution to $\Delta \sigma^{\rm \mysmall NP}_{\pi\pi}$ is proportional to the iso-spin breaking combination $g_V^u - g_V^d$, 
which should be of $\mathcal{O}(1)$ in order to explain $\Delta a_\mu$ in the $m_{Z'} \gtrsim 0.3$ GeV region. 
Therefore, it is natural to expect 
sizeable effects on other iso-spin violating hadronic observables. A relevant example is given by the charged vs.~neutral pion mass squared difference, 
$\Delta m^2 = m^2_{\pi^+} - m^2_{\pi^0}$. Analogously to the QED case, the quadratically divergent $Z'$ loop leads to 
\beq 
\label{eq:Deltam2}
(\Delta m^2)_{Z'} \sim \frac{(g_V^u - g_V^d)^2}{(4\pi)^2} \Lambda^2_\chi \, ,
\eeq
where $\Lambda_\chi  \approx 1$ GeV is the cut-off scale of the chiral theory and we chose $m_{Z'} \ll \Lambda_\chi$.  
Instead, for $m_{Z'} \gg \Lambda_\chi$, the $Z'$ contribution to $\Delta m^2$ decouples as $\Lambda_\chi^2 / m_{Z'}^2 \ll 1$.
In practice, the NP contribution from a light $Z'$ is more reliably obtained by rescaling the SM prediction from lattice QCD~\cite{Frezzotti:2021yuu}
with $(g^u_V - g^d_V)^2 / e^2$.
Then, comparing the SM prediction of $\Delta m^2$ with its experimental value, 
we find the $95\%$ C.L. bound $\abs{g_V^u - g_V^d} \lesssim 0.06$. 

The interplay of the above constraints in the plane $-g_V^e$ vs.~$g_V^u - g_V^d$ is displayed in Fig.~\ref{fig:gegudmZp}
for two representative scenarios where $m_{Z'}=0.1$ and 0.5 GeV. The directions of the arrows indicate the excluded regions 
by the different experimental bounds. Instead, the red band is the region favoured by the explanation of the muon $g$-2 discrepancy. 
From Fig.~\ref{fig:gegudmZp} it is clear that, irrespectively of the $Z'$ mass, there are always at least two independent bounds 
preventing to solve the new muon $g$-2 puzzle.

\Sec{Conclusions.}
The recent lattice QCD result by the BMW collaboration shows a tension with the low-energy 
$e^+e^- \to \text{hadrons}$ data currently used to determine the HVP contribution 
to the muon $g$-2. 
In this Letter we investigated the possibility to solve this tension, 
referred to as the 
\emph{new muon $g$-2 puzzle}, 
invoking NP in the hadronic cross-section. 
A possible way to restore full consistency into the picture is to postulate a \emph{negative} shift in  
$\sigma_{\rm had}$ due to NP.  
We showed that this scenario requires the presence of a light NP mediator that modifies the experimental 
cross-section $\sigma_{\rm had}$. 
However, this non-trivial setup, where NP hides in $e^+e^- \to \text{hadrons}$ data, 
is excluded by a number of experimental constraints. 
Alternative confirmations of the $e^+e^-$ determinations of the HVP contribution 
to the muon $g$-2, based on either additional lattice QCD calculations or direct experimental measurements, as proposed
by the MUonE experiment \cite{CarloniCalame:2015obs,Abbiendi:2016xup,MUonE:LoI},\footnote{It is 
very unlikely that NP contributions will contaminate the MUonE's extraction of $a^{\rm \mysmall HVP}_\mu$~\cite{Dev:2020drf,Masiero:2020vxk}.} 
will hence be crucial to shed light on this intriguing puzzle.

\Sec{Note added.}
After this paper was posted on the arXiv, Ref.~\cite{Darme:2021huc} 
appeared in which the authors 
studied the possibility of reconciling the data driven and the BMWc lattice determinations 
of $a_\mu^{\rm \mysmall HVP}$ by rescaling the KLOE luminosity via a 
NP contribution to Bhabha scattering. 
Hence, contrary to our analysis, in their approach NP does not directly contribute to $\sigma_{\rm had}$.  

\Sec{Acknowledgments.}
We thank M. Fael, A. Keshavarzi, L. Vecchi and G. Venanzoni for helpful discussions. 
This work was partially supported by the European Union's Horizon 2020 research and innovation programme 
under the Marie Sklodowska-Curie grant agreement No 860881-HIDDEN, by 
the research grant ``The Dark Universe: A Synergic Multi-messenger 
Approach'' number 2017X7X85K under the program PRIN 2017 funded by the Ministero dell'Istruzione, Universit\`a 
e 
della 
Ricerca (MIUR) 
and by the INFN Iniziative Specifiche APINE and TAsP.

\vspace{-0.2cm}
\bibliography{bibliography}

\end{document}